\documentclass{PoS}

\title{Top and EW Physics at the LHeC}

\ShortTitle{Top and EW Physics at the LHeC}

\author{\speaker{Zhiqing Zhang}\thanks{for the LHeC Study Group}\\
        Laboratoire de l'Acc\'el\'erateur Lin\'eaire, Univ.\ Paris-Sud 11 et IN2P3/CNRS, France\\
        E-mail: \email{zhang@lal.in2p3.fr}}

\abstract{The LHeC is a proposed upgrade of the LHC to study $ep/eA$ collisions in the TeV regime, by adding a 60\,GeV electron beam through an energy recovery linac. In $ep$, high precision top and electroweak physics can be performed, such as measurements of anomalous top couplings, light quark couplings to the $Z$ boson and the energy dependence of the weak mixing angle $\sin^2\!\theta_W$, for which simulation studies are presented.}

\FullConference{The European Physical Society Conference on High Energy Physics\\
		22--29 July 2015\\
		Vienna, Austria}

\begin{document}

\section{Introduction}
Deep inelastic scattering (DIS) of a point-like lepton beam over a hadron has played a central role in establishing the quark-parton model and QCD starting with fixed target experiments in the late 1960s at SLAC. Later the Gargamelle neutrino-nucleon experiment at CERN has discovered weak neutral currents.
HERA, operated at DESY from 1992 to 2007, was the only $ep$ collider of the world. It has extended the study of the proton structure and quark-gluon interaction dynamics up to a centre-of-mass energy ($\sqrt{s}$) of 320\,GeV corresponding to an extension by two orders of magnitude towards both higher negative four-momentum transfer squared $Q^2$ and lower Bjorken $x$ in comparison with the kinematic region covered by the fixed target experiments.

The LHeC, if realised by adding to the LHC a separate 9\,km racetrack-shaped recirculating superconducting energy recovery linac providing a polarised electron (possibly also positron) beam of 60\,GeV, will be a new $ep$ collider of 1.3\,TeV, running in parallel with the high luminosity phase of the LHC. It has a rich and complementary physics programme to the LHC~\cite{cdr,1211.5102}.  It would enable new precision studies of QCD in general and the precision determination of parton distributions functions (PDFs) in a largely extended kinematic region in particular. It has the potential to reveal new QCD dynamics in an unexplored low $x$ regime where the DGLAP evolution equations may no longer be valid as the latest QCD analysis of the newly combined inclusive neutral and charged current (NC and CC) cross sections at HERA may indicate~\cite{herapdf2}. It would also provide additional and sometimes unique ways for studying top and electroweak (EW) physics as well as Higgs and physics beyond the Standard Model (BSM).

This talk focuses on some of the selected topics on top and EW physics at the LHeC and the writeup is organised as follows. In Sec.~\ref{sec:top}, expected limits on anomalous $Wtb$ couplings from the single top production are presented as an exemple. In Sec.~\ref{sec:ew}, the expected precision determination of light quark couplings to the $Z$ boson and the scale dependence of the weak mixing angle $\sin^2\!\theta_W$ based either on the inclusive NC cross section measurements or on polarisation asymmetries of the NC interactions are shown, followed by a summary in Sec.~\ref{sec:summary}.

\section{Top physics}\label{sec:top}
The top quark is the heaviest particle in the SM, which is believed to be most sensitive to BSM physics. It has not been studied so far by any DIS experiments because of the kinematic limit or too small cross section. Therefore the LHeC will be the first DIS experiment capable to study the directly produced single top quark and top pairs in CC and NC interactions, respectively. 

In the five flavour scheme, the single top-quark production cross section of the $2\to 2$ $t$-channel process $e^-p\to \bar{t}\nu_e+X$ with $\bar{t}\to W^-b$ at $\sqrt{s}=1.3$\,TeV is predicted to be around 2\,pb for un polarised electron beam and increases by a factor of $1+P_e$ with $P_e$ being the degree of the longitudinal polarisation of the beam~\cite{dgkm13}. This cross section value is comparable with that of the Tevatron and smaller by about two orders of magnitude than the LHC at 14\,TeV~\cite{nk15}.
The LHeC has however a much cleaner environment due to the absence of pile-up and underlying events. Therefore this process can be used for many precision measurements within the SM, such as the bottom-quark distribution of the proton, the CKM matrix element $V_{tb}$, the $t$-quark polarisation and the $W$ boson helicity. It can also be used to study deviations from the SM such as the anomalous couplings $Wtb$. In addition, the single top production in the NC protoproduction can be used to study top quark flavour changing neutral current couplings $tq\gamma$ with $q$ being a light quark~\cite{cdr}.

The top pair events are also produced at the LHeC in NC interactions. Even though the rate is lower than at the LHC, the potential for a better measurement of $tt\gamma$ than LHC is good~\cite{bl13} as in the $t\bar{t}$ photoproduction at the LHeC, the highly energetic incoming photon couples only to the $t$ quark so that the cross section depends directly on the $tt\gamma$ vertex, whereas at the LHC the vertex is probed through $t\bar{t}\gamma$ production, where the outing going photon could come from other charged sources such as the top decays products. The DIS regime of $t\bar{t}$ production will also be able to probe the $ttZ$ coupling though with less sensitivity. 
 
A detailed study was performed in~\cite{dgkm13} to evaluate the expected accuracy of measuring the anomalous $Wtb$ couplings at the LHeC based on the single anti-top quark production in $e^-p$ collisions in a model independent way by means of the following effective CP conserving Lagrangian~\cite{dgkm13}
\begin{equation}
{\cal L}_{Wtb}=\frac{g}{\sqrt{2}}\left[W_\mu\bar{t}\gamma^\mu\left(V_{tb}f^L_1P_L+f^R_1P_R\right)b-\frac{1}{2m_W}W_{\mu\nu}\bar{t}\sigma^{\mu\nu}\left(f^L_2P_L+f^R_2P_R\right)b\right] + h.c.
\end{equation}
where $f^L_1(\equiv 1+\Delta f^L_1)$ and $f^R_1$ are left- and right-handed  vector couplings, $f^{L,R}_2$ are left- and right-handed tansor couplings, $W_{\mu\nu}=\partial_\mu W_\nu-\partial_\nu W_\mu$, $P_{L,R}=\frac{1}{2}(1\mp\gamma_5)$ are left- and right-handed projection operators, $\sigma^{\mu\nu}=i/2(\gamma^\mu\gamma^\nu-\gamma^\nu\gamma^\mu)$ and $g=2/\sin\theta_W$. In the SM, $f^L_1\equiv 1$ and $\Delta f^L_1=f^R_1=f^{L,R}_2\equiv 0$.

\begin{figure}[htb]
\begin{center}
\vspace{-2mm}
\includegraphics[width=.495\textwidth]{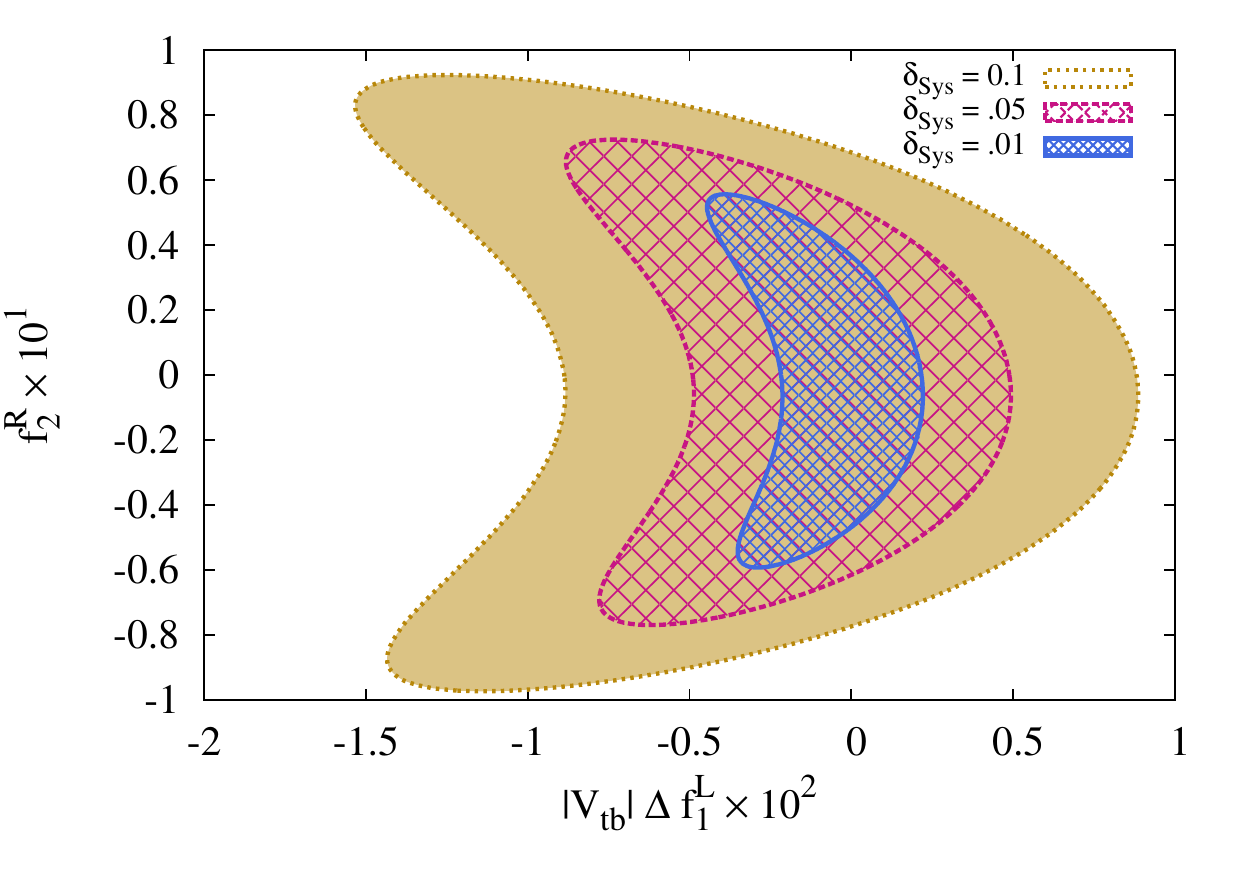}
\includegraphics[width=.495\textwidth]{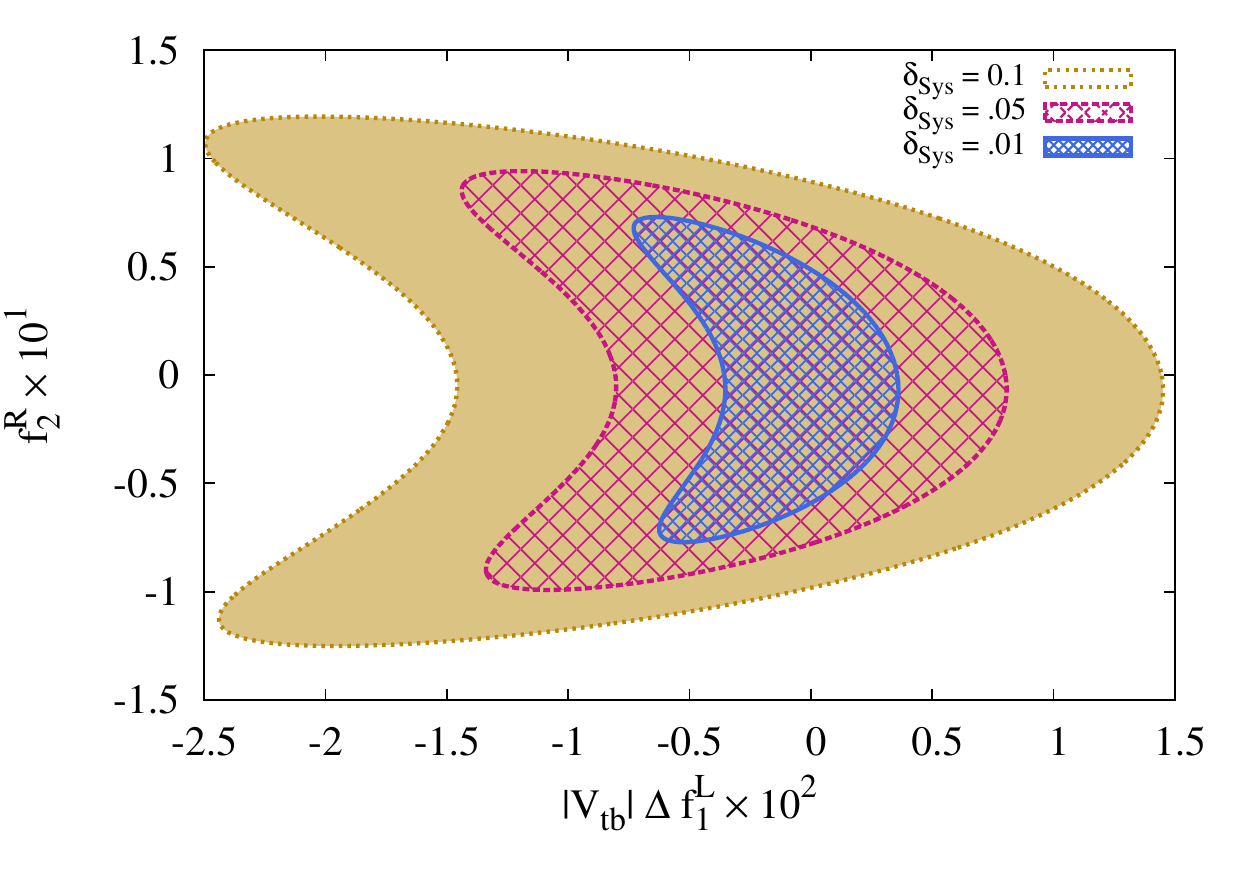}
\end{center}\vspace{-7mm}\caption{Contours at (left) 68\% and (right) 95\% CL on the plane of $|V_{tb}|\Delta^L_1$ and $f^R_2$ for a systematic error of 1\%, 5\% and 10\% on a sample with an integrated luminosity of 100\,fb$^{-1}$ (figures taken from Ref.~\cite{dgkm13}).}
\label{fig:wtb}
\end{figure}

Several analyses were performed using a simulated event sample corresponding to an integrated luminosity of 100\,fb$^{-1}$ for three different systematic uncertainties of 1\%, 5\% and 10\%. One of them was based on a $\chi^2$ analysis using differential distributions of a few relevant kinematic variables in the leptonic and hadronic decay modes, respectively. Contours at 68\% and 95\% confidence level (CL) on two dimensional plane for any coupling combination were presented. One example is shown in Fig.~\ref{fig:wtb}. The corresponding results in comparison with other results from Tevatron, LHC and indirect one from $B$ decays are shown in Table~\ref{tab:wtb}. The conservative LHeC limits are thus competitive with or better than similar results from other determinations.

\begin{table}
\begin{tabular}{l|cccc}
Upper limit {\small (95\% CL)} & $|\Delta f^L_1|$ & $|f^R_1|$ & $|f^L_2|$ & $|f^R_2|$ \\\hline
LHeC~\cite{dgkm13} & $0.005-0.03$ & $0.01-0.1$ & $0.01-0.1$ & $0.01-0.1$ \\
D0~\cite{d0:wtb} & & 0.548 & 0.324 & 0.347 \\
LHC~\cite{lhc:wtb} & $0.03-0.06$ & $0.22-0.34$ & $0.06-008$ & $0.06-0.08$ \\
$B$ decays~\cite{b:wtb} & $[-0.13, 0.03]$ & $[-0.0007, 0.0025]$ & $[-0.0013, 0.0004]$ & $[-0.15, 0.57]$ \\\hline
\end{tabular}
\caption{Comparison of expected upper limits at 95\% CL at the LHeC (100\,fb$^{-1}$, hadronic modes, $\delta_{\rm sy}=0.01-0.1$)~\cite{dgkm13} with the actual limits from D0 (5.4\,fb$^{-1}$, $W$-helicity, single top)~\cite{d0:wtb} and expected limits at the LHC (100\,fb$^{-1}$, $\gamma p\to WtX$)~\cite{lhc:wtb} and $B$ decays (indirect)~\cite{b:wtb}.}
\label{tab:wtb}
\end{table}

\section{EW physics}\label{sec:ew}
Inclusive NC and CC DIS interactions are two main processes which can be measured at the LHeC with high precision providing primary source not only for precision QCD studies but also for EW physics. Three examples are briefly presented in this section.

%
The first example concerns a precision measurement of vector and axial-vector weak NC couplings of the $Z$ boson to light quarks $v_q$ and $a_q$. They were determined together with PDFs in a combined EW and QCD analysis of simulated inclusive NC and CC cross section data samples following Ref.~\cite{h1ew}. This is possible since the double differential NC cross section $\frac{{\rm d}^2\sigma_{\rm NC}}{{\rm d}x{\rm d}Q^2}$ in e.g.\ $e^-p$ collisions may be expressed in terms of three structure functions as $\frac{2\pi\alpha^2}{xQ^2}\left[Y_+\tilde{F}_2+Y_-\tilde{F}_3-y^2\tilde{F}_L\right]$, where $\alpha$ is the electromagnetic fine structure constant and $Y_\pm=1\pm (1-y)^2$ with $y=Q^2/(xs)$ being the electron inelasticity. The generalised structure $\tilde{F}_2$ can be decomposed as $F_2+P_ea_e\kappa_ZF_2^{\gamma Z}+a^2_e\kappa^2_ZF^Z_2$ corresponding to $\gamma$ exchange, $\gamma Z$ interference and $Z$ exchange contributions. In this expression, $\kappa_Z^{-1}=\frac{2\sqrt{2}\pi\alpha}{G_FM^2_Z}\frac{Q^2+M^2_Z}{Q^2}$ and $a_e$ is the axial-vector coupling of the electron (due to the smallness of the vector coupling $v_e$, terms proportional to $v_e$ have been  omitted). Similarly $x\tilde{F}_3=-a_e\kappa_ZxF_3^{\gamma Z}-P_ea^2_e\kappa^2_ZxF_3^{\gamma Z}$. These different structure functions can be further expressed in terms of PDFs $q, \bar{q}$ and the light quark couplings $v_q$ and $a_q$ as $\left[ F_2, F_2^{\gamma Z}, F_2^Z\right]=x\sum_q\left[e^2_q, 2e_qv_q, v^2_q+a^2_q\right]\{q+\bar{q}\}$ and $\left[xF_3^{\gamma Z}, xF_3^Z\right]=2x\sum_q\left[e_qa_q, v_aa_q\right]\{q-\bar{q}\}$, where $e_q$ being the electronic charge of quark $q$. The longitudinal structure function $\tilde{F}_L$ does not contribute at LO. The CC cross section is independent of these  couplings but its inclusion in the fit helps to constrain the PDFs.

\begin{figure}[htb]
\begin{center}
\vspace{-2mm}
\includegraphics[width=.495\textwidth]{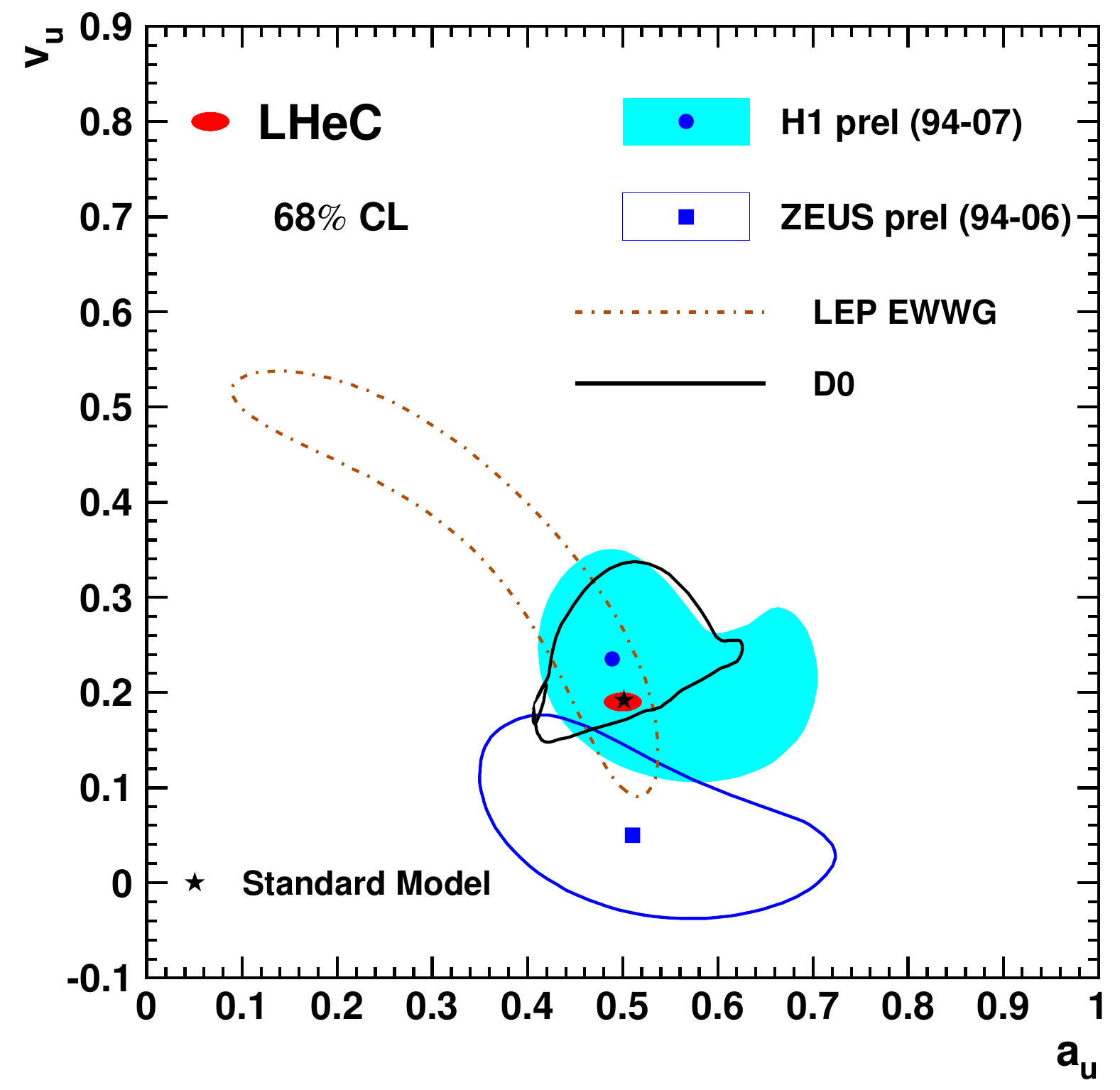}
\includegraphics[width=.495\textwidth]{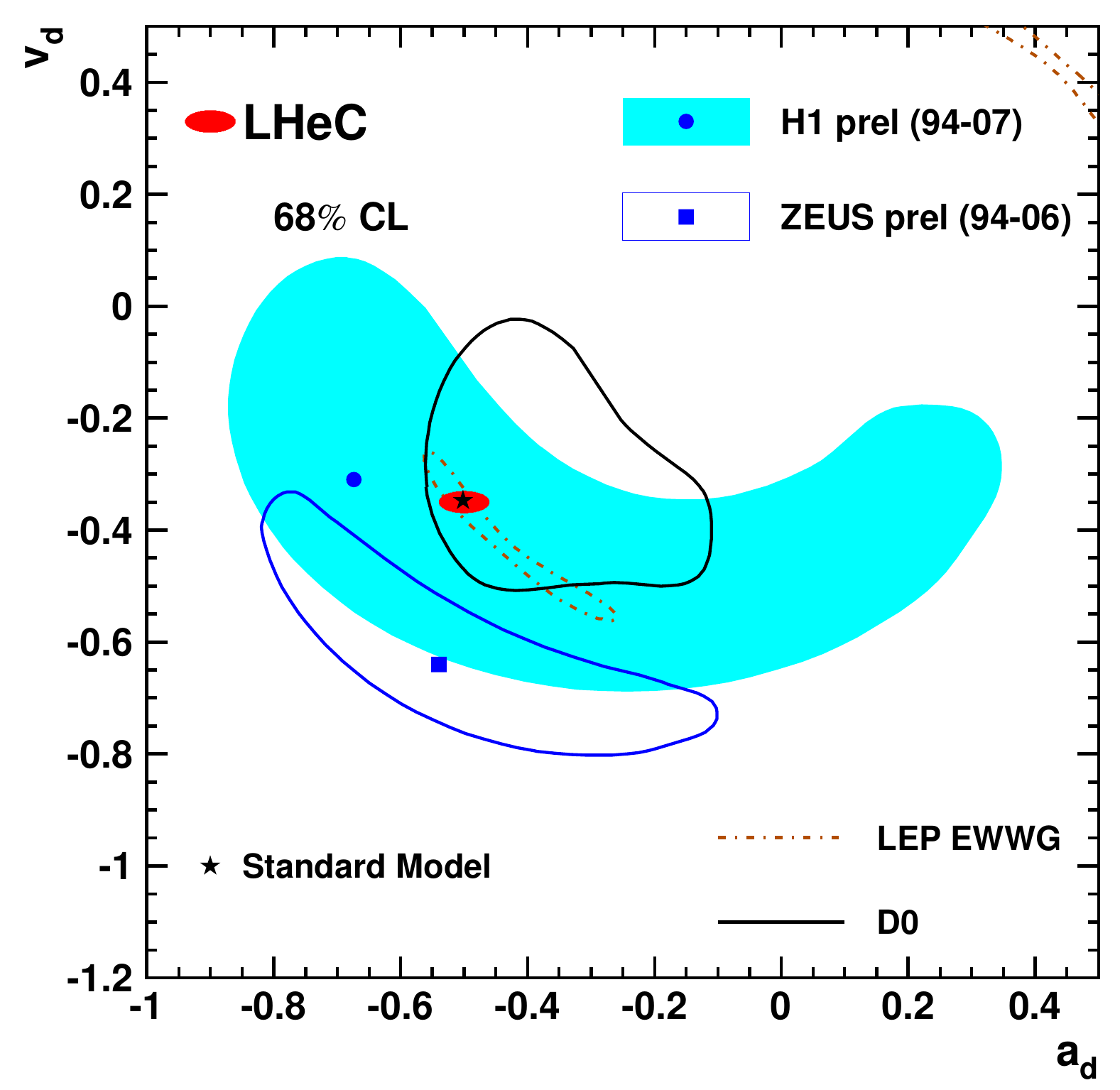}
\end{center}\vspace{-7mm}\caption{Determination of the vector and axial-vector weak neutral current couplings of the light quarks by LEP~\cite{lep_couplings}, D0~\cite{d0_couplings}, H1~\cite{h1_couplings}  and ZEUS~\cite{zeus_couplings}, compared with the simulated prospects for the LHeC~\cite{cdr}.}
\label{fig:couplings}
\end{figure}

In Ref.~\cite{cdr}, different scenarios were considered. The results of one of these, corresponding to $e^\pm$ beams of 50\,GeV with a longitudinal polarisation of 40\% colliding with a proton beam of 7\,TeV for an integrated luminosity of 1\,fb$^{-1}$ per beam, are shown in Fig.~\ref{fig:couplings} in comparison with similar determinations from other experiments. The expected precision at the LHeC is indeed much better and any significant deviation from the SM expectations can thus be observed with this analysis.  

\begin{figure}[htb]
\vspace{-2mm}
\includegraphics[width=.575\textwidth]{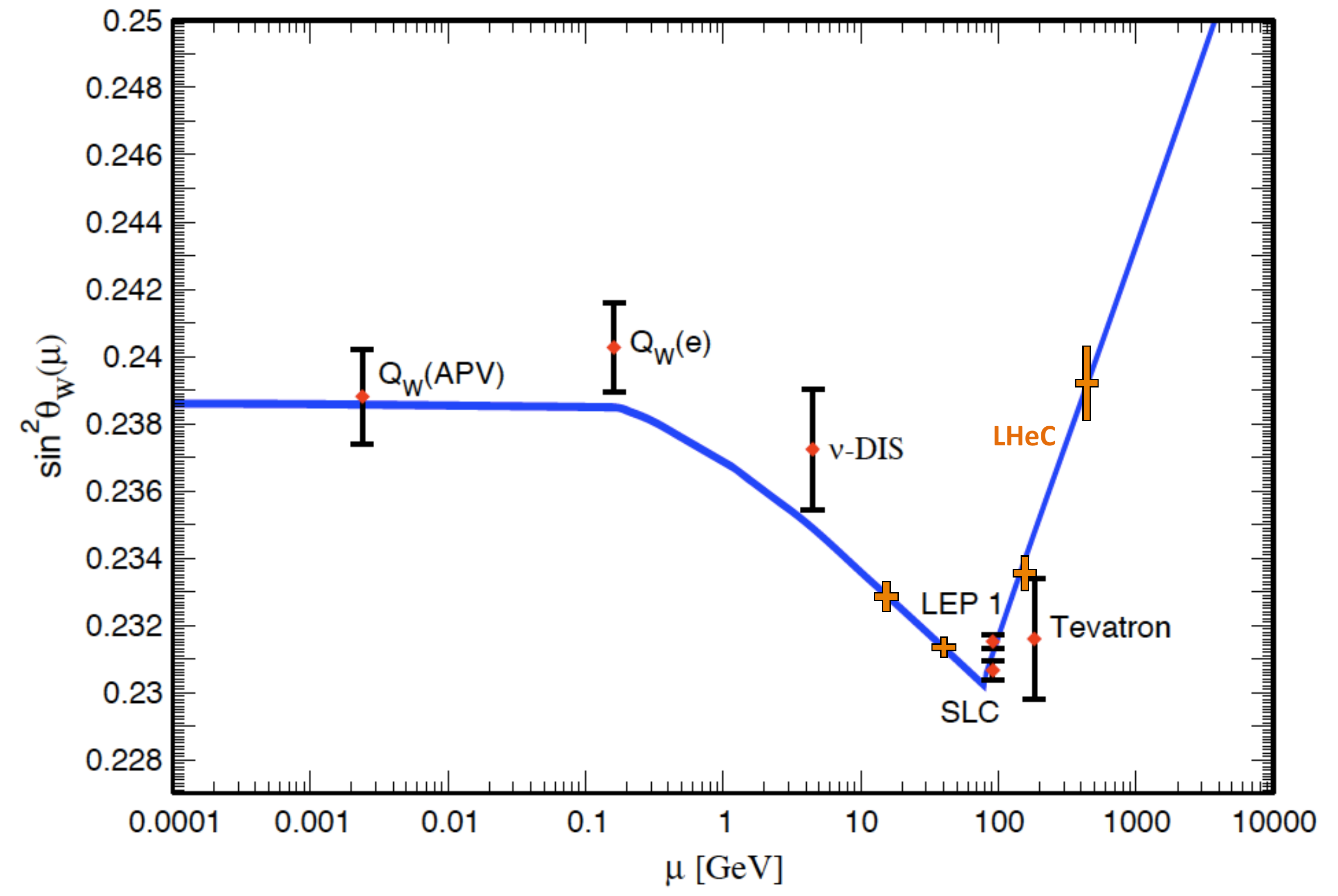}
\includegraphics[width=.43\textwidth]{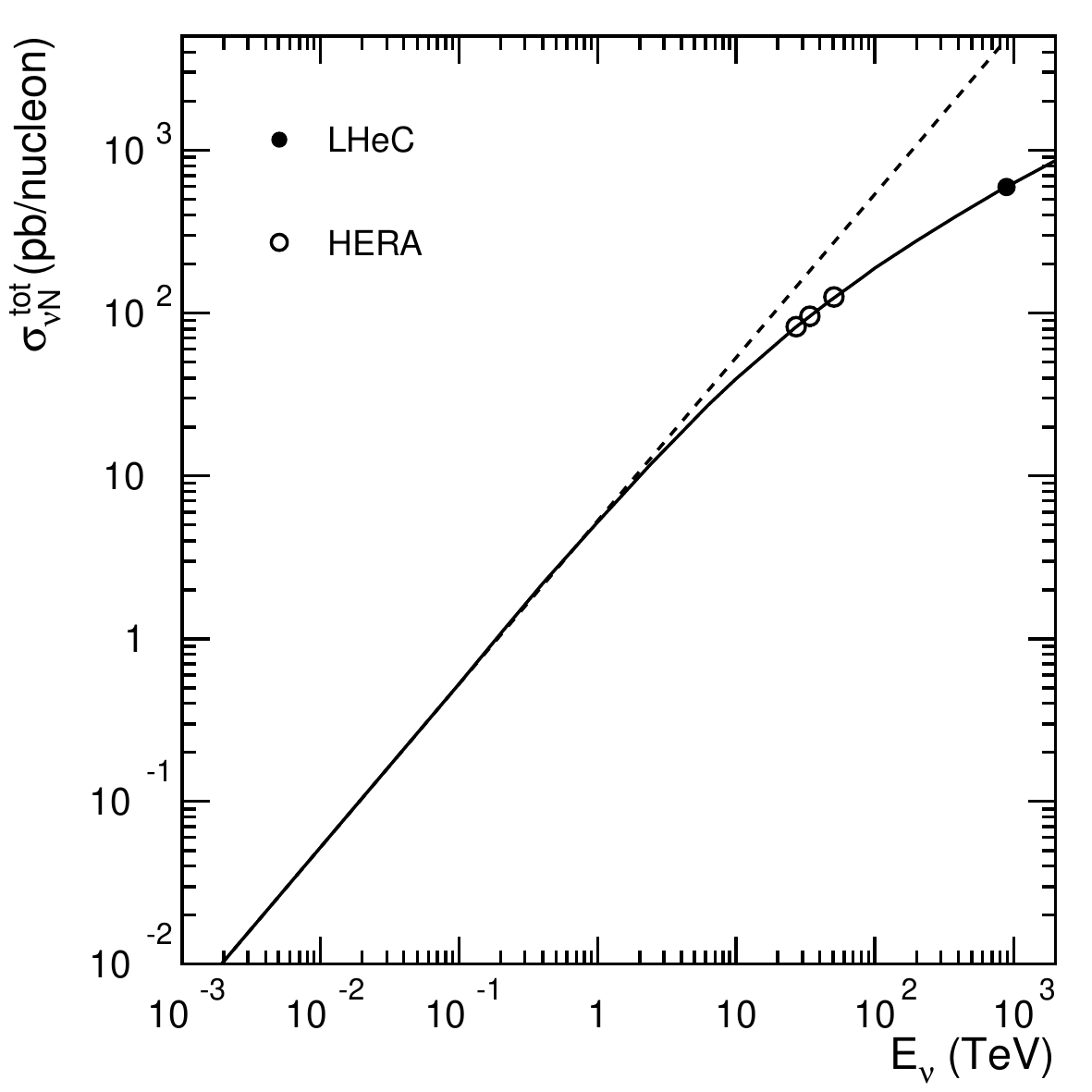}
\caption{Left: dependence of the weak mixing angle on the energy scale $\mu$ from \cite{cdr}. Right:  energy dependence of the $\nu N$ cross section. The open points up to 50\,TeV correspond to the expected precision of the HERA measurements and the solid point corresponds to an expected measurement at the LHeC. The full line represents the predicted cross section including the $W$ propagator while the dashed line is a linear extrapolation from low energy measurements.}
\label{fig:wma_cctot}
\end{figure}

The second example is on the scale dependence of the weak mixing angle $\sin^2\!\theta_W$ obtained from a projected measurement of the polarisation asymmetry $A^-=\frac{\sigma^-_{\rm NC}(P_R)-\sigma^-_{\rm NC}(P_L)}{\sigma^-_{\rm NC}(P_R)+\sigma^-_{\rm NC}(P_L)}\simeq \frac{\kappa_Za_e(P_L-P_R)}{2}\frac{F_2^{\gamma Z}}{F_2}$ assuming a left-handed ($P_L$) or right-handed ($P_R$) polarisation of 80\% and an integrated luminosity of 10\,fb$^{-1}$ per polarisation state~\cite{cdr}. The results are compared in Fig.~\ref{fig:wma_cctot} (left) with other determinations at different energies. The LHeC measurements are precise and cover a large energy range. It should be mentioned that the NC and CC cross section ratio is also sensitive to $\sin^2\!\theta_W$, provided that the PDFs related uncertainty is under control~\cite{cdr}.  

The third example is on a measurement of the CC total cross section (Fig.~\ref{fig:wma_cctot} (right)). The LHeC measurement together with those from HERA illustrates in a spectacular way the impact of the propagator mass of the $W$ boson on the CC cross section. The dependence on the polarisation of the CC total cross section can further be used to set a stringent lower mass limit on a right-handed $W$ boson following Ref.~\cite{h1cc}.

\section{Summary}\label{sec:summary}
A few selected examples of the LHeC measurements from top and EW sectors clearly demonstrate that the realisation of the LHeC can greatly enhance the physics programme and discovery potential of the LHC in a complementary manner. 
Indeed the LHeC is also considered as the next machine for studying the Higgs boson
  and a luminosity upgrade by a factor of 10 reaching $10^{34}\,{\rm cm}^{-2}s^{-1}$ is under study~\cite{max}, compared to
  the Conceptional Design Report~\cite{cdr}. 
More studies are desirable from both theoretical and experimental communities, which will certainly reveal even larger potential of the LHeC.

\end{document}